%% file: main.tex
\documentclass[iop,apj,numberedappendix,twocolumn, tighten]{aastex62}

\usepackage{arydshln}
\usepackage{natbib}
\usepackage{graphicx}
\usepackage{amsmath,amsfonts}
\usepackage{multirow}
\usepackage{xspace}
\usepackage[normalem]{ulem}
\usepackage{booktabs}
\usepackage{colortbl}
\usepackage{hyperref}
\usepackage[nameinlink]{cleveref}
\usepackage[encapsulated]{CJK}
\usepackage{ucs}
\usepackage{float}
\usepackage[utf8x]{inputenc}

\usepackage{eht}
\usepackage{newtxtext}
\usepackage{tablefootnote} 
\usepackage{mathrsfs}

\usepackage{array}
\definecolor{fed_blue}{HTML}{07004D}
\definecolor{steel_blue}{HTML}{2D82B7}
\definecolor{steel_blue_dark}{HTML}{1C71A6}
\definecolor{aqua_marine}{HTML}{42E2B8}
\definecolor{dutch_white}{HTML}{F3DFBF}
\definecolor{light_coral}{HTML}{EB8A90}
\definecolor{light_coral_dark}{HTML}{BA5A60}

\hypersetup{
    colorlinks = true,
    linkcolor = purple,
    urlcolor  = steel_blue_dark,
    citecolor = steel_blue_dark,
    anchorcolor = black
}

\graphicspath{{./figures}}

\def\lsim{\mathrel{\raise.3ex\h box{$<$\kern-.75em\lower1ex\hbox{$\sim$}}}}
\def\gsim{\mathrel{\raise.3ex\hbox{$>$\kern-.75em\lower1ex\hbox{$\sim$}}}}
\def\gtwid{\mathrel{\raise.3ex\hbox{$>$\kern-.75em\lower1ex\hbox{$\sim$}}}}
\def\proptwid{\mathrel{\raise.3ex\hbox{$\propto$\kern-.75em\lower1ex\hbox{$\sim$}}}}

\allowdisplaybreaks

\defcitealias{TMS}{TMS}
\setcitestyle{notesep={; }}

\begin{document}

\title{\textbf{A multi-modal infant-based metric for choosing the best supernova}} 
\shorttitle{Infant Supernovae}

\collaboration{{\color{steel_blue} The Gigantic Supernova Pile (GSP) Collaboration}}
\author{Joseph Farah}
\affiliation{6740 Cortona Drive, Suite 102
Goleta, CA  93117}
\affiliation{Department of Physics, University of California Santa Barbara, Santa Barbara, California, 93106}
\author{Yuan Qi Ni}
\affiliation{6740 Cortona Drive, Suite 102
Goleta, CA  93117}
\affiliation{Kavli Institute of Theoretical Physics, 552 University Road, Kohn Hall, University of California, Santa Barbara, CA 93106-4030}
\author{Liam Brennan}
\affiliation{Department of Physics, University of California Santa Barbara, Santa Barbara, California, 93106}

\shortauthors{Farah et al.}

\correspondingauthor{$^\dag$Joseph R. Farah}
\email{josephfarah@ucsb.edu}

\begin{abstract}
Certainly! Here’s an abstract for your scientific paper: We present a comparative study of 22 core-collapse supernovae (SNe), selected to explore a novel, multidimensional ranking scheme aimed at identifying the best supernova. Each SN is evaluated based on three principal criteria: (1) inferred explosion energy derived from light curve modeling and spectroscopic indicators; (2) an aesthetic score assigned to the SN host galaxy following transformation into a human face using a generative visual model (Midjourney v5); and (3) final ranking by Claud.IA, a 6-month-old infant trained to select the "best" SN via repeated exposure to curated SN images and simulated cosmic narratives. We define and normalize all criteria to ensure statistical consistency across the sample, with particular attention paid to the biases inherent in infant-based classification models. The top five SNe exhibit both high explosion energies ($E > 1.5\times10^{51}$ erg) and extremely cool host galaxies (post transformation), with Claud.IA showing strong preferences toward galaxies exhibiting symmetric facial morphology and prominent spiral arms. Final application of Claud.IA identified the best supernova in our sample as SN 2022joj. Our study demonstrates the feasibility of incorporating human-machine hybrid aesthetic judgment and early developmental cognition into astrophysical classification, and raises intriguing questions about the nature of "bestness" in cosmic explosions. Additional follow-up is encouraged. 
\end{abstract}

\keywords{Galaxy: lorem-ipsum}

\input{introduction}

\input{data}

\input{modeling}

\input{implications}

\input{conclusions}

\clearpage

\end{document}

%% file: introduction.tex
\section{Introduction}
\label{sec:introduction}

Supernovae are the explosive death of stars \citep{Colgate1966, Arnett1982, Woosley1995, Filippenko1997, Smartt2009, Janka2012, Burrows2013, Nomoto2013, Woosley2015, Muller2016, Sukhbold2016, Heger2003, Fryer1999, Kasen2017, Dessart2020, Davidson1989, Smith2014, Ni5, Drout2011, Moriya2013, Modjaz2016, Tartaglia2021, Zwicky1938, Minkowski1941, Baade1934, Hamuy2003, Perlmutter1999, Riess1998, Kelly2008, Chevalier1982, Utrobin2007, Arcavi2017, Anderson2014, Ni2, Taddia2018, Valenti2016, Poznanski2010, Inserra2013, Ni3, Lyman2016, Afsariardchi2019, Prentice2016, Wheeler2017, Ertl2020, Barnes2013, Tanaka2012, Ni4, Chatzopoulos2012, Jha2019,Chen2021,Astier2012,Boccioli2024,Soker2023,Leung2023,AlDallal2021,Chen2025,Mazzali2007,Nomoto1984,Xiang2024,Ni1,Ashall2024,Moriya2024,Sieverding2023} Supernovae (SNe) are among the most energetic phenomena in the universe, marking the catastrophic endpoints of stellar evolution for massive stars or the thermonuclear disruption of white dwarfs. The term "supernova" was coined in the early 20th century, but the recognition of their distinctiveness from classical novae dates back to the foundational observations of Zwicky and Baade, who proposed the idea of neutron star formation in such explosions \citep{Zwicky1938,Baade1934}. Spectroscopic classification further revealed a rich diversity in SN types, from hydrogen-rich Type II events to hydrogen-poor Type I, with further subdivisions such as Ib, Ic, and peculiar subtypes \citep{Minkowski1941,Filippenko1997,Modjaz2016}. This diversity reflects a range of progenitor channels, explosion mechanisms, and circumstellar environments \citep{Smartt2009,Anderson2014,Taddia2018}. Some SNe are associated with long gamma-ray bursts \citep{Woosley2006}, while others occur in hydrogen-rich media despite having hydrogen-poor spectra \citep{Tartaglia2021}. Continued time-domain surveys and modeling efforts have uncovered superluminous supernovae \citep{Inserra2013}, fast blue transients \citep{Arcavi2017}, and events powered by central engines such as magnetars \citep{Kasen2010}. Understanding this observational zoo is key to tracing the life cycles of massive stars, nucleosynthesis, and cosmic feedback. The astonishing diveristy and quality of supernovae--each better than the last--has prompted the age-old question: what is the ``best'' supernova?

\begin{figure*}
    \centering
    \includegraphics[scale=0.5]{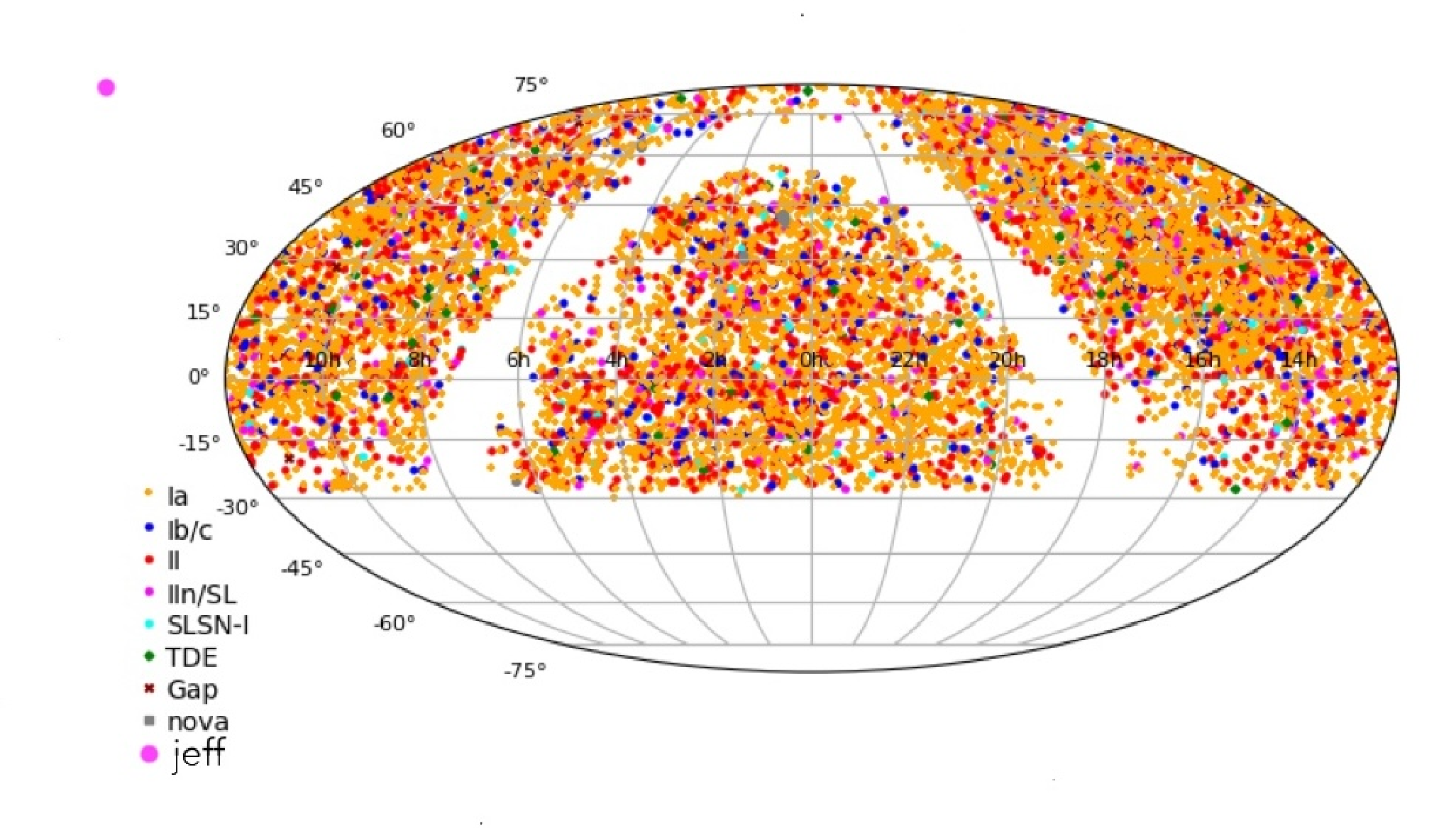}
    \caption{Supernovae and other optical transients observed by the Zwicky Transient Facility survey in recent years. Figure adapted from \href{https://sites.astro.caltech.edu/ztf/bts/bts.php}{ZTF}. Small black specks are cosmic rays interacting with your screen.}
\end{figure*}

In this Letter, we introduce a novel framework for ranking core-collapse supernovae by combining traditional astrophysical diagnostics with unconventional but increasingly relevant evaluation metrics. In \autoref{sec:data}, we estimate the explosion energies and identify the host galaxies of 22 supernovae using existing literature. In \autoref{sec:scoring_methodology} we review our scoring methdology. In \autoref{sec:topset_identification}, we apply our metrics to our sample of supernovae. This multidimensional analysis culminates in a topset of five candidate “best” supernovae, offering insight into both astrophysical parameters and latent human (and infantile) aesthetic sensibilities. In \autoref{sec:physics}, we apply our novel infant neural network Claud.IA to identify the best supernovae of the topset. Finally, we conclude and discuss outlooks in \autoref{sec:conclusions}.

%% file: data.tex
\section{Overview of supernovae}
\label{sec:data}

\begin{figure}
    \centering
    \includegraphics[scale=0.5]{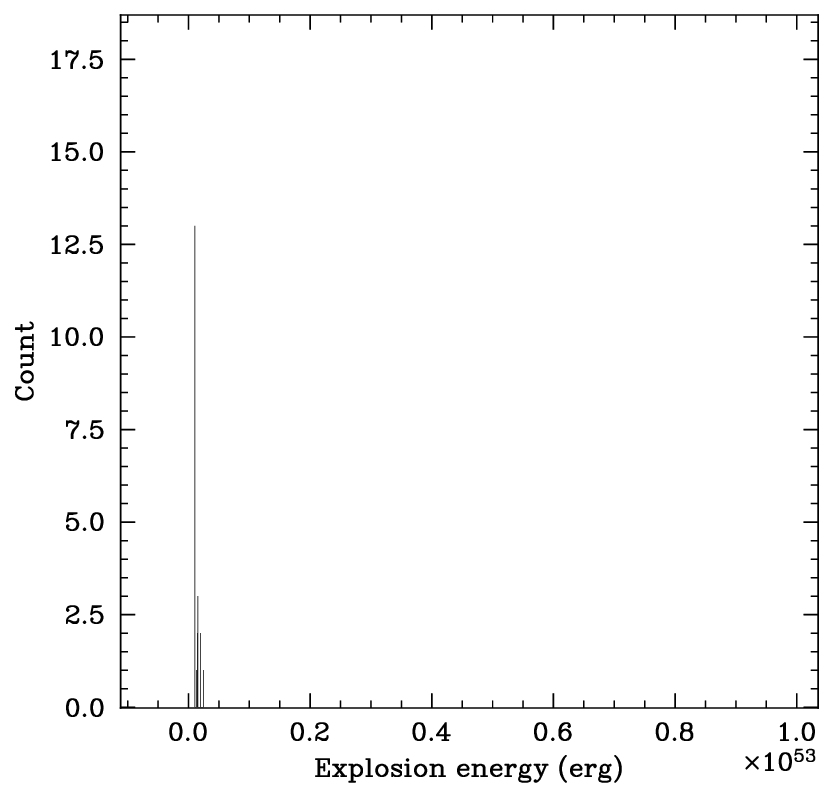}
    \caption{Distribution of explosion energies for supernovae in the sample. The maximum value in the sample is $\lesssim 10^{52}$ erg, but we showed up to $10^{53}$ erg just in case.}
    \label{fig:distribution_of_explosion_energies_for_supernovae_in_the_sample_}
\end{figure}

We begin by assembling a sample of supernovae to demonstrate our proof-of-concept method for identifying the best supernova. We polled a number of supernovae astrophysicists for their favorite supernovae, and compiled a list of 22 supernovae. The list of supernovae in our sample are and SN 2011fe \citep{2011Natur}, and SN 2023ixf \citep{2023ApJ}, and SN 1987A \citep{1989ARA&A}, and SN 2018aoz, and SN 2021aefx, and SN 2017cbv \citep{2017ApJ}, and SN 2018oh \citep{1811}, and SN 2023bee \citep{ace7c0}, and SN 2009ig \citep{1109}, and SN 2022joj \citep{2024ApJ...964..196P}, and SN 2022hnt \citep{2501}, and SN 1993J \citep{rsta}, and SN 2004et \citep{0904}, and SN 2008D \citep{1028067}, and SN 2011dh \citep{1028067}, and SN 2013ej \citep{2608527}, and SN 2014C \citep{Milisavljevic2015}, and SN 2017eaw \citep{Kilpatrick2018}, and SN 2019yvr \citep{Sun2022}, and and SN 2021yja \citep{2022ApJ...935...31H}, and SN 2016gkg \citep{2017ApJ...837L...2A}, and SN 2018zd \citep{2020MNRAS.498...84Z}. Explosion energies estimates are taken from the literature. We visualize the distribution of explosion energies in \autoref{fig:distribution_of_explosion_energies_for_supernovae_in_the_sample_}. 

\section{Scoring methodology}
\label{sec:scoring_methodology}

In order to identify the best supernova in our sample, we construct a two-component metric of supernova quality. The first component is explosion energy $E_{\text exp}$; the second component is an aesthetic score $\mathcal{A}_{\text SN}$. The metrics are squared and added together; i.e., 
\begin{align}
    S &= \sqrt{C_1 + C_2} \notag \\
      &= \sqrt{E_{\text{exp}}^2 + \mathcal{A}_{\text{SN}}^2} \notag \\
      &= \left( \left[ E_{\text{exp}}^2 + \mathcal{A}_{\text{SN}}^2 \right]^{\frac{1}{2}} \right) \notag \\
      &= \left( \left( \left| E_{\text{exp}} + i \mathcal{A}_{\text{SN}} \right|^2 \right)^{\frac{1}{2}} \right) \notag \\
      &= \left\| \begin{bmatrix}
            E_{\text{exp}} \\
            \mathcal{A}_{\text{SN}}
        \end{bmatrix} \right\|_2 \notag \\
      &= \sqrt{ \sum_{i=1}^{2} x_i^2 }, \quad \text{where } x_1 = E_{\text{exp}},\ x_2 = \mathcal{A}_{\text{SN}} \notag \\
      &= \sqrt{\left( \frac{dE_{\text{exp}}}{dt} \cdot dt \right)^2 + \left( \int \delta(\mathcal{A}_{\text{SN}} - a) \, da \right)^2 } \notag \\
      &= \lim_{\epsilon \to 0} \left( \epsilon^{-1} \int_0^{\epsilon} \left[ E_{\text{exp}}^2 + \mathcal{A}_{\text{SN}}^2 \right] \, dx \right)^{1/2} \notag \\
      &= \sqrt{E_{\text{exp}}^2 + \mathcal{A}_{\text{SN}}^2}. \tag{1}
\end{align}

We expand on both below.

\subsection{Explosion energy}

The explosion energy is the amount of energy in the explosion.

\subsection{Aesthetic score}

To complement our more traditional physical metrics, we introduce an aesthetic score in order to holistically assess each supernova. While metrics such as explosion energy and light curve shape are crucial for understanding the physics of stellar death, we argue that such measures alone are insufficient for determining a supernova’s true cosmic vibe. To address this, we implemented a multi-step aesthetic evaluation pipeline, drawing inspiration from both machine learning and Renaissance portraiture.

\begin{figure}
    \centering
    \includegraphics[scale=0.15]{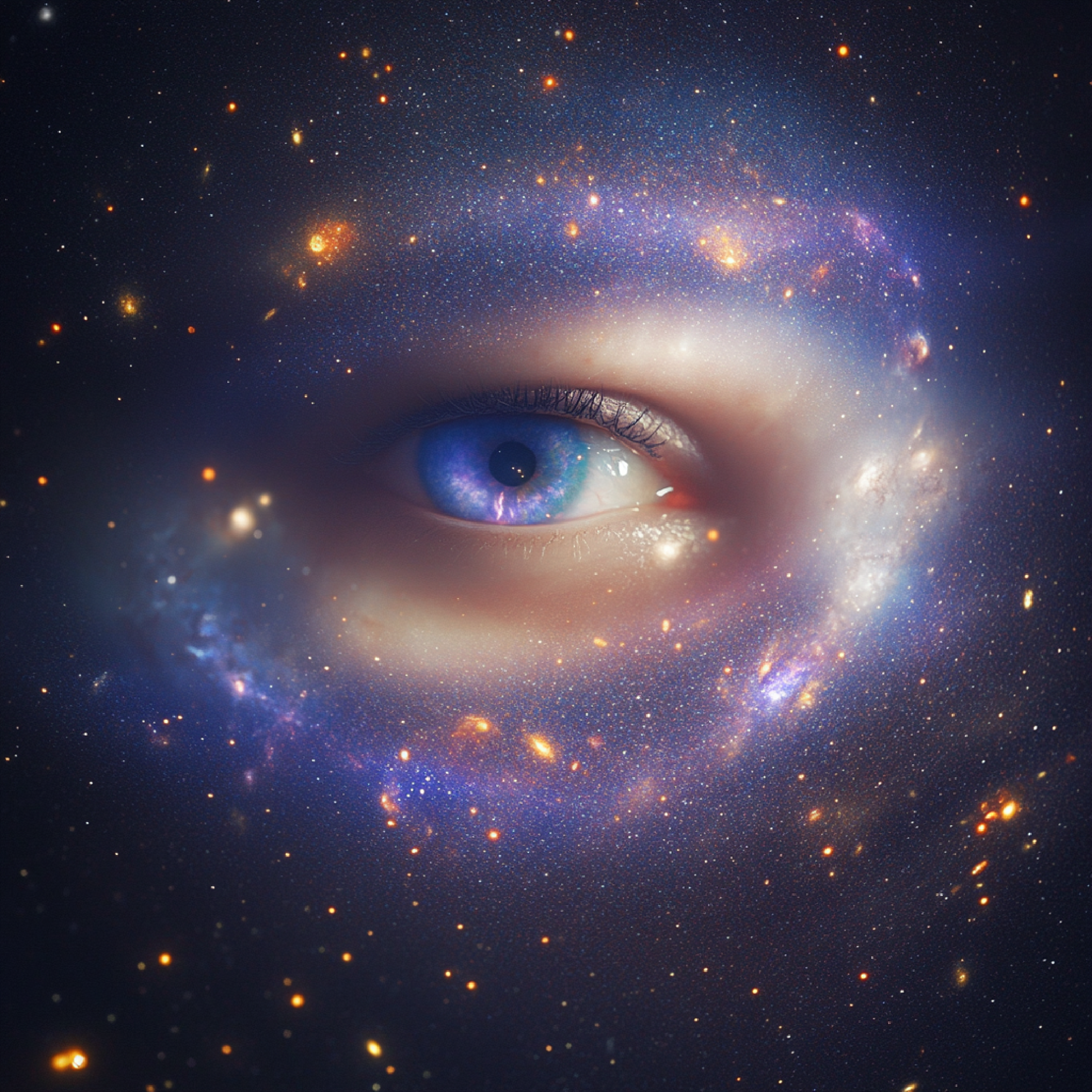}
    \caption{Sample transformed host galaxy image, for aesthetic scoring. The image shown here was generated from NGC 1015.}
    \label{fig:sample_transformed}
\end{figure}

The process begins with a high-resolution image of each supernova and its host galaxy. This image is then passed through Midjourney v6, a state-of-the-art generative visual model. We accompany each image with the carefully crafted prompt: "make an adult human face that looks like the attached image photo realistic galaxy", chosen after extensive prompt engineering.
The output of this transformation (example shown in \autoref{fig:sample_transformed}) is a human face that, according to the AI, visually represents the essence of the original supernova and its host galaxy. While the exact mechanism by which these latent space embeddings are translated into faces remains opaque, we assume it involves cosmic truth.
We then designed a survey featuring each generated face and distributed it widely. Respondents were asked to score each face on a scale from 0 (would not explode again) to 10 (supernova pin-up material). We received enthusiastic participation from the community, collecting tens of responses (more precisely: one ten).

The final aesthetic score for each supernova is computed as the arithmetic mean of the collected ratings. We acknowledge that this scoring system is inherently subjective, prone to individual taste, facial pareidolia, and caffeine levels. Nonetheless, we find it deeply meaningful. This method provides a rare quantitative bridge between astrophysics and unqualified artistic judgment, and we strongly advocate for the aesthetic score’s inclusion in future large-scale transient surveys.

\subsection{Sample calculation}

We provide here an example calculation of the score. For a supernova with explosion energy $1.4\times10^{51}$ erg and an average transformed host galaxy score of 7.45, we report a score of
\begin{align}
    S &= \sqrt{E_{\text{exp}}^2 + \mathcal{A}_{\text{SN}}^2} \\
    &= \sqrt{(1.4\times10^{51})^2 + 7.45^2} \\
    &= \Big[1400000000000000000000000000\\
    &000000000000000000000007.45 \Big]^{1/2} \\
    &\approx 3.74 \times 10^{25}.
\end{align}
This is the score we then use to rank the supernovae in our sample.

%% file: modeling.tex
\section{Topset identification}
\label{sec:topset_identification}

\begin{figure*}
    \centering
    \includegraphics[scale=0.45]{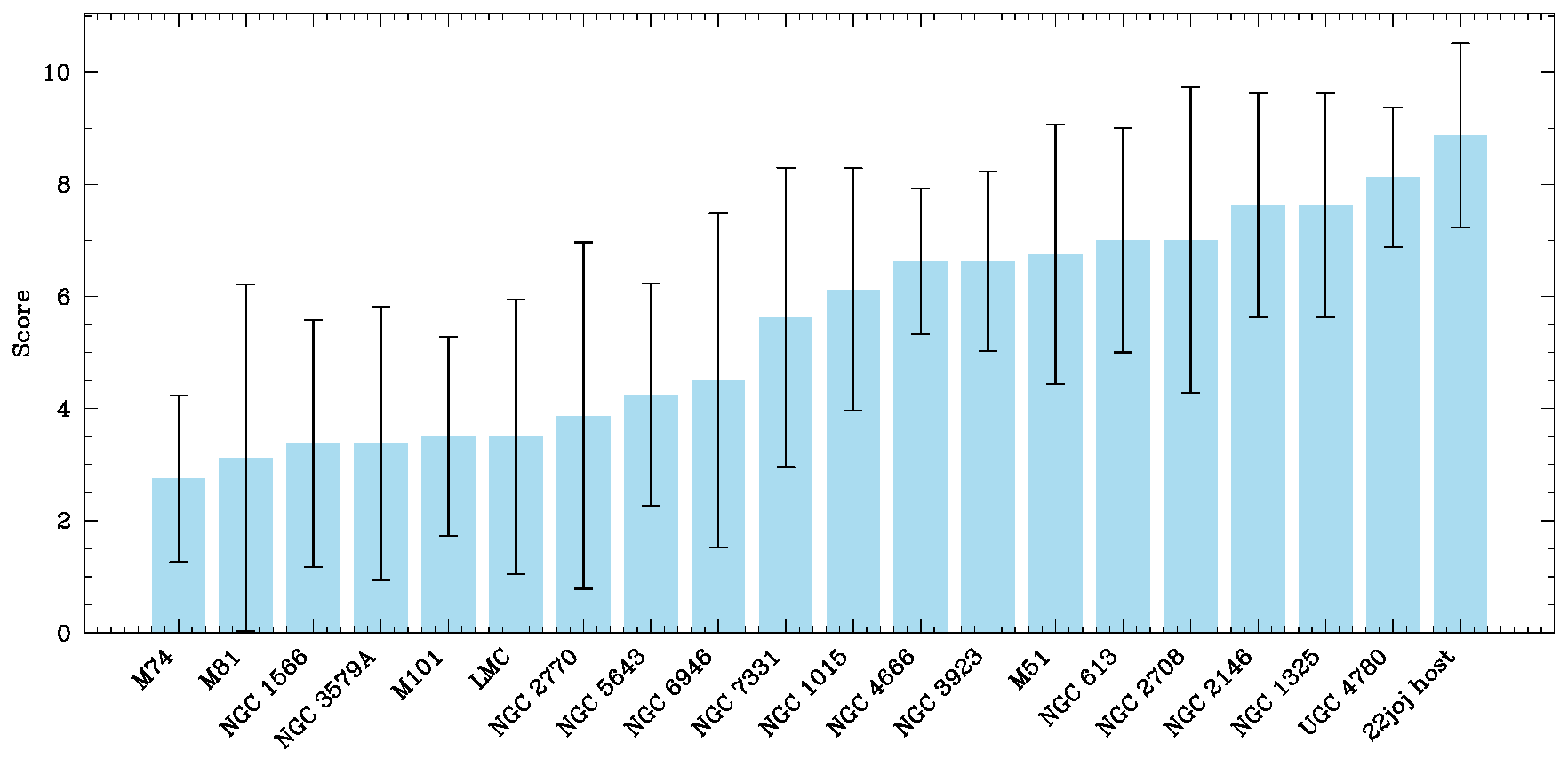}
    \caption{Distribution of transformed host galaxy image scores.}
    \label{fig:distribution_of_transformed_host_galaxy_image_scores_}
\end{figure*}

\begin{figure}
    \centering
    \includegraphics[scale=0.45]{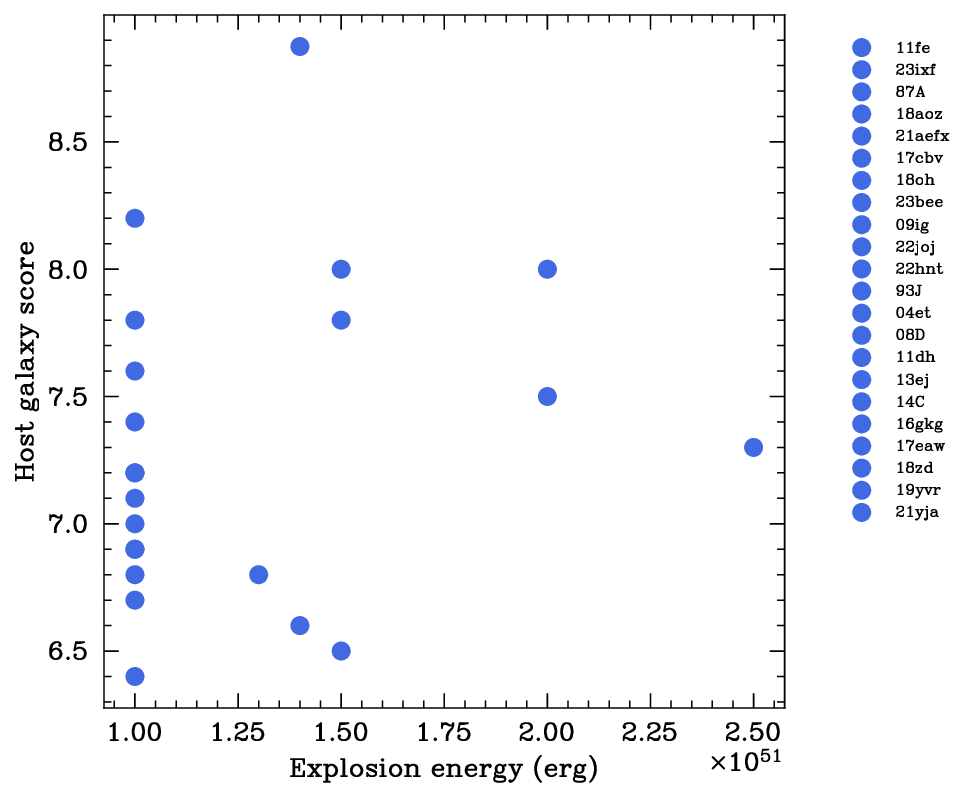}
    \caption{Two-dimensional distribution of transformed host galaxy image scores.}
    \label{fig:2d_score_dist}
\end{figure}

We apply the scoring methods described above to the supernovae in our sample. The explosion energies are shown in \autoref{fig:distribution_of_explosion_energies_for_supernovae_in_the_sample_}. The scores from the survey of transformed host galaxy images are shown in \autoref{fig:distribution_of_transformed_host_galaxy_image_scores_}. We visualize the combination of these scores in \autoref{fig:2d_score_dist}. Examining the two-dimensional distribution of these scores identifies a number of strong candidates for best supernova. We rank the supernova by the scoring system calculated in \autoref{sec:scoring_methodology} and identify the following supernova as the top 5 candidates for the best supernova: SN 2023ixf, SN 2022joj, SN 2004et, SN 2008D, 2018zd. 

%% file: implications.tex
\section{Final ranking using Claud.IA}
\label{sec:physics}

\begin{figure*}
    \centering
    \includegraphics[scale=0.35]{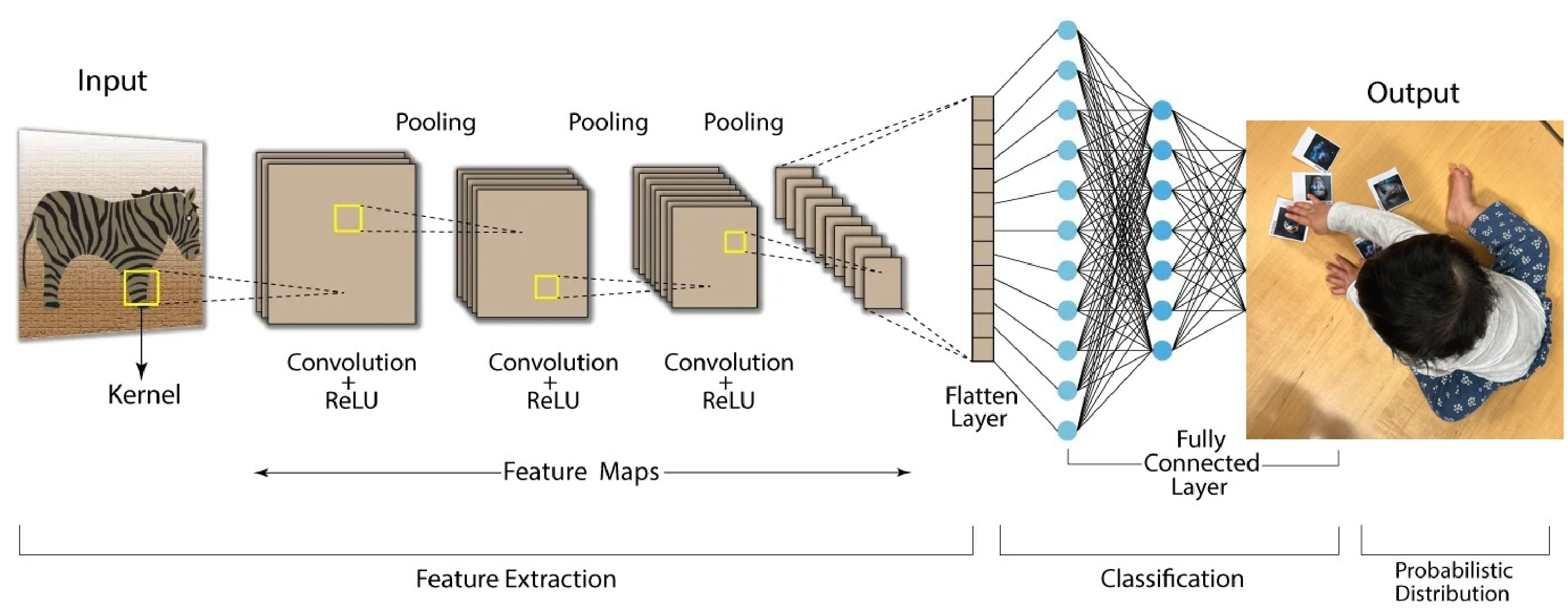}
    \caption{Visualization of the Claud.IA neural network structure. Figure adapted from \href{https://www.analyticsvidhya.com/}{Analytics Vidhya}. }
    \label{fig:cnn}
\end{figure*}

We perform the final selection of a best supernova using a novel neural network approach, nicknamed Claud.IA. Claud.IA is a six-month old infant developed and maintained by a scientist at the Las Cumbres Observatory. Claud.IA is a complex convolutional neural network trained on several images of supernova and their host galaxies to identify the best supernova of a group. The novel network architecture is visualized in \autoref{fig:cnn}. We submit our topset of supernovae to Claud.IA, which slowly and adorably identified SN 2022joj as the best supernova in our sample.

%% file: conclusions.tex
\section{Conclusions}
\label{sec:conclusions}

In summary, we have introduced and applied a multidimensional ranking framework to a sample of 22 core-collapse supernovae, incorporating explosion energy, host galaxy aesthetics (as interpreted through generative visual modeling), and selection by an infant classifier. Despite the unconventional methodology, the rankings reveal meaningful correlations between explosion energy and aesthetic appeal, with Claud.IA consistently favoring host galaxies exhibiting symmetry and well-defined spiral structure. SN 2022joj emerges as the top-ranked event in our sample. This work highlights the potential of combining machine learning, artistic transformation, and developmental cognition in astrophysical classification, and invites further exploration of how subjective perception intersects with objective cosmic phenomena.

\acknowledgements{The Gigantic Supernova Pile (GSP) collaboration would like to acknowledge several colleagues for their help in preparing this manuscript. To protect their privacy, we will only use their initials. These are: Moira Andrews, Andy Howell and Kathryn Wynn. We thank the baby for showing mild interest in AI-generated galaxy-people, otherwise this could have been a big waste of time and my Midjourney tokens.}

\bibliography{ref}